# Spin-waves in 2D honeycomb lattice *XXZ*-type van der Waals antiferromagnet CoPS$_3$

Chaebin Kim[1,2,3], Jaehong Jeong[2,3*], Takatsugu Masuda[4], Shinichiro Asai[4], Shinichi Itoh[5], Heung-Sik Kim[6], Andrew Wildes[7], and Je-Geun Park[1,2,3$]

[1]Center for Quantum Materials, Seoul National University, Seoul 08826, Korea
[2]Center for Correlated Electron Systems, Institute for Basic Science, Seoul 08826, Korea
[3]Department of Physics and Astronomy, Seoul National University, Seoul 08826, Korea
[4]Institute for Solid State Physics, The University of Tokyo, Chiba 277-8581, Japan
[5]Institute of Materials Structure Science, High Energy Accelerator Research Organization, Tsukuba 305-0801, Japan
[6]Department of Physics, Kangwon National University, Chuncheon 24311, Korea
[7]Institut Laue-Langevin, CS 20156, 38042 Grenoble Cédex 9, France

\* hoho4@snu.ac.kr
$ jgpark10@snu.ac.kr

**Abstract**

The magnetic excitations in CoPS$_3$, a two-dimensional van der Waals (vdW) antiferromagnet with spin *S*=3/2 on a honeycomb lattice, has been measured using powder inelastic neutron scattering. Clear dispersive spin waves are observed with a large spin gap of ~13 meV. The magnon spectra were fitted using an *XXZ*-type $J_1$-$J_2$-$J_3$ Heisenberg Hamiltonian with a single-ion anisotropy assuming no magnetic exchange between the honeycomb layers. The best-fit parameters show ferromagnetic exchange $J_1$=-2.08 meV and $J_2$=-0.26 meV for the nearest and second-nearest neighbors and a sizeable antiferromagnetic exchange $J_3$=4.21 meV for the third-nearest neighbor with the strong easy-axis anisotropy *K*=-2.06 meV. The suitable fitting could only be achieved by the anisotropic *XXZ*-type Hamiltonian, in which the exchange interaction for the out-of-plane component is smaller than that for the in-plane one by a ratio $\alpha=J_z/J_x$=0.6. Moreover, the absence of spin-orbit exciton around 30 meV indicates that Co$^{2+}$ ions in CoPS$_3$ have a *S*=3/2 state rather than a spin-orbital entangled $J_{eff}$=1/2 ground state. Our result directly shows that CoPS$_3$ is an experimental realization of the *XXZ* model with a honeycomb lattice in 2D vdW magnets.

## 1. Introduction

Magnetism in two-dimensional (2D) systems has long been one of the most exciting topics in the condensed matter physics. Notably, they are the ideal playground for novel phenomena in 2D magnetic systems, such as Berezinskii-Thouless-Kosterlitz transition of the *XY* model [1,2] and the Mermin-Wagner theorem for the Heisenberg model [3]. The recent introduction of magnetic van der Waals (vdW) materials opens enormous opportunities to examine the low-dimensional magnetism in real materials [4]. Since they are coupled by a weak vdW force along the *c*-axis, it is easy to drive to the 2D limit of the bulk magnetic properties by exfoliation. These systems have so far shown various magnetic properties, including antiferromagnetic *TM*PS$_3$ (*TM*=transition metal) [5], ferromagnetic honeycomb Cr$X_3$ (*X*=Cl, Br, I) [6–8], Cr$_2$Ge$_2$Te$_6$ [9], VI$_3$ [10], and antiferromagnetic triangular *TMX$_2$* family [11].

The *TM*PS$_3$ (*TM*=Mn, Fe, Co, Ni) family has attracted special interests in the community as a class of antiferromagnetic 2D vdW materials [5,12–15]. The transition metal (TM) ions with 2+ covalency in this family form a layered honeycomb lattice with the sulfur ligand ions. They all have the same monoclinic structure with a space group *C*2/*m*, where layers on the *ab*-plane are coupled by a weak vdW force along the *c*-axis [13]. Since the magnetic structure and exchange interactions depend on the TM elements, they provide an excellent playground to experimentally validate the theory of magnetic dynamics in low dimensions [16–24]. For example, FePS$_3$ is described as an ideal Ising magnet [16–18,25], whereas MnPS$_3$ and NiPS$_3$ are the examples of the Heisenberg model [16,19–22,26,27]. Among them, NiPS$_3$ is known to have a magnetic order close to an *XY*-type [23,26]. The thickness dependence of their physical and magnetic properties has been also extensively investigated [14,23,25,27].

By comparison, CoPS$_3$ is less studied among *TM*PS$_3$ due to the difficulty in synthesizing high purity samples [24]. It is known to have an antiferromagnetic order below $T_N$=120 K and shows a zig-zag magnetic structure with the propagation vector **Q**$_m$=(0,1,0), as shown in Figure 1. The spins at the Co sites are aligned along the *a*-axis with a small canting to the *c*-axis [24]. The magnetic susceptibility shows a difference between **H**//*ab* and **H**//*c* in the paramagnetic region, which is the evidence for an *XY*-like anisotropy [24]. It implies that CoPS$_3$ has anisotropic magnetic interactions depending on the magnetic moment direction. Of further interest, the high-spin $d^7$ configuration in Co$^{2+}$ compounds was recently speculated to host a spin-orbital entangled $J_{\text{eff}}$=1/2 state with a Kitaev interaction via the edge-sharing network as in the low-spin $d^5$ case for several 5$d$ materials [28,29]. If found correct, CoPS$_3$ would then offer an exciting new opportunity of examining the Kitaev physics combined with strong correlation of Coulomb *U*. Therefore, it is urgent to investigate this possibility experimentally.

To understand this complex magnetism in CoPS$_3$, one needs to examine the underlying spin Hamiltonian in detail. Although the magnetic structure provides some information about the spin Hamiltonian, it is not sufficient enough to justify the exact spin Hamiltonian chosen. For instance, the *XY* model ($J_x=J_y$, $J_z=0$) and the isotropic Heisenberg model ($J_x=J_y=J_z$) with easy-plane anisotropy can have the same magnetic ground state. Yet, their magnon spectra cannot be, a priori, the same for the Hamiltonian with different symmetries. Inelastic neutron scattering (INS) is the most powerful technique to measure the spin dynamics, and so to determine the type of spin Hamiltonian and the strength of exchange interactions. INS can also be used to verify the existence of the spin-orbital entangled $J_{eff}$=1/2 ground state by the observation of excitation from $J_{eff}$=1/2 to $J_{eff}$=3/2 states, which is expected to occur around 20-30 meV and used as the characteristic signature of the $J_{eff}$=1/2 ground state in cobalt compounds [30–32].

In this paper, our INS experiment and analysis show that CoPS$_3$ has an *S*=3/2 state instead of a $J_{eff}$=1/2 state. And we report the easy-plane *XXZ*-type spin Hamiltonian and exchange parameters of CoPS$_3$. We also compare the exchange parameters and the effect of single-ion anisotropy of CoPS$_3$ with those for another *TM*PS$_3$ family [17–19,21].

## 2. Experimental Methods

Powder samples of CoPS$_3$ were synthesized by a solid-state reaction of the pure elements. Stoichiometric quantities of cobalt, phosphorus, and sulfur were placed in a quartz ampoule under an Ar atmosphere. The total mass of the ingredients was 2 g, and the purity of the elements was 99.99% or better. The ampoule was then evacuated, sealed under 5 torr of argon environment, and heated in a tube furnace. The temperature was raised to 530°C in 6 hours and held there for 2 days. We performed an inelastic neutron scattering experiment using a high-resolution chopper spectrometer HRC at J-PARC facility, Japan [33]. Taking advantage of the repetition rate multiplication method with the chopper frequency of 200 Hz, we measured the INS data at 8, 35, 60, 85, 110, and 200 K with the fixed incident neutron energies of $E_i$=71.3 and 40.3 meV. Measurements at room temperature were also carried out for the background subtraction with each incident energy.

## 3. Data analysis & modeling

The measured INS data were reduced and binned using the MSlice program in the DAVE suite [34]. To obtain pure magnon dispersions without phonon contamination, we subtracted the room-temperature data from the data at low temperatures after applying the Bose factor. The measured magnetic excitations were modeled and fitted using the linear spin-wave theory. We used the *XXZ*-type (anisotropic Heisenberg) Hamiltonian with a single-ion anisotropy:

$$H = \sum_{n=1}^{3} J_n \sum_{<i,j>_n} [S_i^x S_j^x + S_i^y S_j^y + \alpha S_i^z S_j^z] + K \sum_i (\hat{x} \cdot S_i)^2, \quad (1)$$

where $\alpha \in [0,1]$ is the spin anisotropy parameter that spans from the *XY* model ($\alpha$=0) to isotropic Heisenberg model ($\alpha$=1), $K$ is the strength of the single-ion anisotropy, and $J_n$ is the exchange interaction up to the third nearest neighbors. Since the inter-layer interaction is presumably negligible for a weak vdW force, we ignore the interlayer coupling in our analysis.

In our work, we examined both the isotropic Heisenberg model and the anisotropic *XXZ* model specifically. Using a Holstein-Primakoff transformation and Fourier transform, one can rewrite the spin Hamiltonian with a quadratic form of magnon operators:

$$H = \sum_Q X^\dagger \mathcal{H} X + N E_{MF}, \quad (2)$$

where $E_{MF}$ is the mean-field ground state energy per spin, and $N$ is the total number of spin sites. For the isotropic Heisenberg model, we define the magnon creation and annihilation operators as $X^\dagger = [\hat{b}_{Q,1}^\dagger, \hat{b}_{Q,2}^\dagger, \hat{b}_{-Q,3}, \hat{b}_{-Q,4}]$, where the number in subscription denotes each sublattice, *i.e.*, $\hat{b}_{Q,i}^\dagger$ ($\hat{b}_{Q,i}$) creates (annihilates) a magnon on the *i*th sublattice. Then the Hamiltonian matrix in Eq. (2) becomes

$$\mathcal{H} = \begin{pmatrix} A & B^* & C & D^* \\ B & A & D & C \\ C & D^* & A & B^* \\ D & C & B & A \end{pmatrix}, \quad (3)$$

where

$$A = S(-J_1 + 4J_2 \cos^2(\pi h) + 3J_3 - 2K),$$

$$B = S J_1 \exp\left(\pi i \left(h + \frac{k}{3}\right)\right)(1 + e^{-2\pi i h})/2,$$

$$C = 2SJ_2 (\cos(\pi(h+k)) + \cos(\pi(h-k))),$$

$$D = Se^{2\pi ik}\left(J_1 + J_3\left(\exp\left(\frac{4\pi ik}{3}\right) + 2\cos(2\pi h)\right)\right)/2.$$

Here *h* and *k* are the components of the wavevector **Q**=(*h*,*k*) in units of the reciprocal lattice vector.

For the *XXZ* model, there are more magnon coupling terms between magnetic sublattices due to the anisotropic nature of the magnetic interaction. Thus, we need to use the operator basis of full 8 terms, $X^\dagger = [\hat{b}^\dagger_{Q,1}, \hat{b}^\dagger_{Q,2}, \hat{b}^\dagger_{Q,3}, \hat{b}^\dagger_{Q,4}, \hat{b}_{-Q,1}, \hat{b}_{-Q,2}, \hat{b}_{-Q,3}, \hat{b}_{-Q,4}]$. The Hamiltonian matrix can be written as

$$\mathcal{H} = \begin{pmatrix} A & B^* & C & D^* & E & F^* & G & H^* \\ B & A & D & C & F & E & H & G \\ C & D^* & A & B^* & G & H^* & E & F^* \\ D & C & B & A & H & G & F & E \\ E & F^* & G & H^* & A & B^* & C & D^* \\ F & E & H & G & B & A & D & C \\ G & H^* & E & F^* & C & D^* & A & B^* \\ H & G & F & E & D & C & B & A \end{pmatrix}, \quad (4)$$

where

$$A = S(-J_1 + J_2(2 + (\alpha + 1)\cos(2\pi h)) + 3J_3 - 2K),$$

$$B = S(\alpha + 1)J_2 \exp\left(\pi i\left(h + \frac{k}{3}\right)\right)(1 + e^{-2\pi ih})/2$$

$$C = S(\alpha - 1)J_2(\cos(\pi(h+k)) + \cos(\pi(h-k))),$$

$$D = S(\alpha - 1)e^{2\pi ik}\left(J_1 + J_3\left(\exp\left(\frac{4\pi ik}{3}\right) + 2\cos(2\pi h)\right)\right)/2$$

$$E = S(\alpha - 1)J_2 \cos(2\pi h),$$

$$F = S(\alpha - 1)\exp\left(\pi i\left(h + \frac{k}{3}\right)\right)(1 + e^{-2\pi ih})/2$$

$$G = S(\alpha + 1)J_2(\cos(\pi(h+k)) + \cos(\pi(h-k))),$$

$$H = S(\alpha + 1)e^{2\pi ik}\left(J_1 + J_3\left(\exp\left(\frac{4\pi ik}{3}\right) + 2\cos(2\pi h)\right)\right)/2$$

The Hamiltonian matrices in Eq. (3) and (4) can be diagonalized to determine the eigenvectors, which are then used to calculate the magnetic dynamic structure factor *S*(**Q**,*E*). Once convoluted with the proper instrument resolution function, it then gives the partial differential neutron cross-section, which corresponds to the measured neutron intensity. The spin-wave dispersion and powder averaged neutron cross-section were calculated by the SpinW package [35]. For our analysis, we selected the data up to *Q*=2.5 Å$^{-1}$, excluding the elastic scattering below *E*=8 meV. The difference between the simulation results and the measured data was minimized by using the particle swarm optimization algorithm [36].

## 4. Result

Fig. 2 shows the spin waves taken at 8 K with an incident neutron energy $E_i$=71.3 meV, together with representative linear spin-wave theory calculations. As one can see, the measured data show dispersive magnons with a large spin gap of ~13 meV. It shows another gap around 25 meV so that there are two magnon modes, with one being a flat upper band and another a lower dispersive one. Fig. 2a and 2c show the simulated powder averaged neutron cross-section using the best-fit parameters with the convolution of instrumental resolution of 3 meV for the *XXZ*-type and isotropic Heisenberg models, respectively. The best-fit parameters for the isotropic Heisenberg model give ferromagnetic exchange interactions for the first and second-nearest neighbors, $J_1$=-2 meV and $J_2$=-0.65 meV, and a significant antiferromagnetic third nearest neighbor exchange $J_3$=3.51 meV. A strong easy-axis single-ion anisotropy $K$=-3.62 meV is necessary to open the large gap, as observed in the experiments. When we tried the *XXZ* model, we found that while the sign of exchange parameters is the same as the isotropic Heisenberg model, their actual values are slightly changed: $J_1$=-2.08, $J_2$=-0.26, $J_3$=4.21 and $K$=-2.06 meV with an easy-plane spin anisotropy factor $\alpha=J_z/J_x$ is 0.6.

In Fig. 3, we plot the constant-$Q$ and constant-$E$ cuts integrated over the range denoted in Fig. 2 with vertical and horizontal white boxes to present a detailed comparison between the two models. As clearly shown in Fig. 3a and 3b, the isotropic Heisenberg model is not consistent with the low energy gap and the extra gap-like spectra at 24-27 meV. Notably, the best fit single-ion anisotropy $K$=-3.62 meV for the isotropic model overestimates the low energy gap. As shown in Fig. 3c, the isotropic Heisenberg model gives very little intensity at the bottom of observed magnon dispersion for the range of energy from 13 to 16 meV around $Q$=1.6 and 2.2 Å$^{-1}$. When we used the isotropic Heisenberg model in Fig. 3d, it produces a significant intensity between the lower and upper magnon bands, unlike the observed gap-like feature in the range of energy from 24 to 27 meV.

The temperature dependence of magnon dispersion is also depicted in Fig. 4. The phonon contamination is subtracted from the data by the same method used for the analysis of the 8 K data. In Fig. 4a-f, the intensity of magnon dispersions slowly decreases as temperature increases before magnons are dramatically collapsed near the Néel temperature $T_N$=120 K. But a significant portion of the original spectral weights is found to survive above the $T_N$ due to over-damped spin waves from critical fluctuations. To demonstrate this point better, we plot the temperature dependence of the integrated intensity through the overall range of $Q$=0.3-4 Å$^{-1}$ in Figs. 4g and 4h.

## 5. Discussion

The magnetic structure of CoPS$_3$ that we used in the spin-wave calculation is the zig-zag order with spin moments aligned to the crystallographic *a*-axis. We accordingly set the direction of single-ion anisotropy to the *a*-axis. It has been suggested from the detailed analysis of the magnetic susceptibility with a crystal field splitting that the easy-axis anisotropy lies in the *ac*-plane and aligns along the *a*-axis [24]. However, the actual magnetic moment direction measured by neutron diffraction has a small canting by 10.5° to the *c*-axis [24]. To take into account the reported structure, we also fitted the data with the canted magnetic structure as well, but the best-fit parameters and the agreement factor did not change much. For the isotropic Heisenberg model, it is easily expected because the direction of single-ion anisotropy only forces the moment direction but does not alter the magnon spectra. In the case of the *XXZ* model, we had to set a larger tilting in the single-ion anisotropy to stabilize the reported canting angle due to the easy-plane spin anisotropy. However, by a similar reason as the isotropic Heisenberg model, it makes no meaningful difference.

The planar type spin anisotropy in CoPS$_3$ can be interpreted as originating from the trigonally distorted environment of Co$^{2+}$ ion in the octahedron. Under the trigonal elongation crystal field, the $t_{2g}$ state split into the upper singlet $a_{1g}$ and lower doublet $e_g^\pi$ states. The lifted orbital degeneracy leads to anisotropic interactions between the in-plane and out-of-plane direction of the magnetic moment [37]. There are also other quasi-2D antiferromagnetic honeycomb compounds, which show similar *XXZ*-type interactions such as BaCo$_2$(*A*O$_4$)$_2$ (*A*=P, As) with the strong spin anisotropy factor $\alpha$=0.4 [38,39]. Both compounds also have the trigonal elongation in the local CoO$_6$ octahedra to the *c*-axis. It suggests that the origin of the easy-plane spin anisotropy has a close relationship with the orbital splitting by trigonal elongation.

Fig. 5. shows the expected spin waves of CoPS$_3$ along with some high symmetry directions in the Brillouin zone based on our best-fit parameter set for both isotropic Heisenberg and the *XXZ* model. Since the spin anisotropy lowers the symmetry of Hamiltonian, the magnon branches of the *XXZ* model get further split as compared with those of the isotropic Heisenberg model. Also, all these magnons in the *XXZ* model are separated along with the trajectories, and this separation makes the extra spin gap at 24 - 27 meV. Moreover, at the Brillouin zone center and points C, the number of non-degenerated magnon branches is doubled in the *XXZ* model. We observe that the lowest non-degenerated magnon branch at 14meV is well-matched with the measured low-energy gap.

In Table 1, we summarize all the experimentally measured magnetic exchange parameters for the $TM$PS$_3$ family, with the parameters defined in a self-consistent way with Eq. (1), including our result for CoPS$_3$ [17,19,21]. It is noticeable that despite the same atomic structure and a similar typical antiferromagnetic order, the parameters are found to depend strongly on the magnetic TM ion. It is also noteworthy that the second-nearest neighbor exchange interaction is commonly small for all $TM$PS$_3$, which are supported by density functional theory as well [40].

We would now like to examine our best-fit parameters by considering magnetic properties. Using a mean-field theory, we can estimate the magnetic ordering temperature $T_N$ and the Curie-Weiss temperature $\Theta_{CW}$ as follows:

$$k_B \Theta_{CW} = -\frac{1}{3} S(S+1)(3J_1 + 6J_2 + 3J_3),$$

$$k_B T_N = -\frac{1}{3} S(S+1)(J_1 - 2J_2 - 3J_3). \cdots (5)$$

The obtained temperatures using our best-fit parameters are found to be $\Theta_{CW}$=-43.5 K and $T_N$=188.5 K, in good agreement with the reported value of $\Theta_{CW}$=-35.4 K [24] and $T_N$=120 K. The slightly more significant difference in the estimated $T_N$ is probably due to critical fluctuations as reported in NiPS$_3$ and MnPS$_3$ [16,19]. This set of parameters also stabilizes the said zig-zag magnetic order [41]. It therefore provides further confidence to our conclusion that the parameters obtained are reasonable and agree with the magnetic properties of the bulk sample. Another interesting point to note is that there is no sign of other higher-energy excitations in our data, such as flat-like spin-orbit excitons for the $J_{eff}$=1/2 state (see Fig. 4a-f). It is in stark contrast with other Co-based compounds, which exhibit excitations around 20-30 meV due to magnetic excitons [30–32]. The absence of such excitations directly implies that CoPS$_3$ has a spin $S$=3/2 state rather than a spin-orbital entangled $J_{eff}$=1/2 ground state.

Finally, there are several similarities in magnetism between CoPS$_3$ and NiPS$_3$. Both compounds show the zig-zag antiferromagnetic order with the sizeable third-nearest neighbor interaction $J_3$. For NiPS$_3$, the $J_3$ value is much larger than $J_1$ as $J_3/J_1$=3-4, which is the largest among the $TM$PS$_3$ family. In our analysis for CoPS$_3$, $J_3$ is also as large as $J_3/J_1$=2. Those large $J_3$ values are theoretically explained by a robust super-super-exchange interaction between the $e_g$ orbitals of third-neighboring Ni ions through the sulfur ligand ions [42,43]. In NiPS$_3$, two holes reside in the $e_g$ orbitals, whereas another hole exists in $t_{2g}$ in CoPS$_3$. This $t_{2g}$ orbital contribution to the super-super-exchange can reduce the strength of antiferromagnetic $J_3$. A recent study suggests that NiPS$_3$ also has an easy-plane spin anisotropy [23]. Although the strength of single-ion anisotropy is quite different due to the difference of electronic configuration, it is interesting that the different orbitals in $CoPS_3$ and $NiPS_3$ possess a similar spin anisotropy.

## 6. Conclusion

We successfully determined the magnetic exchange parameters and the single-ion anisotropy of $CoPS_3$ using powder inelastic neutron scattering. We found that $Co^{2+}$ in the $CoPS_3$ has a spin $S$=3/2 state, not a spin-orbital entangled $J_{eff}$=1/2 ground state. So, the observed magnon spectra were fitted well by the anisotropic *XXZ*-type $J_1$-$J_2$-$J_3$ Heisenberg Hamiltonian with a sizable easy-axis single-ion anisotropy of the $S$=3/2 state. The best fit result shows the ferromagnetic $J_1$ and $J_2$, and antiferromagnetic $J_3$ exchange interactions for the first-, second- and third-nearest neighbors in the *XXZ* model. We observed the large spin gap of ~13 meV, which requires a sizable single-ion anisotropy of $K$=-2.06 meV. The analysis shows that $CoPS_3$ has the *XXZ*-type interaction in a honeycomb lattice with an easy-plane spin anisotropy α=0.6. Therefor our experiment and theoretical analysis put $CoPS_3$ as another useful example of *XXZ*-type antiferromagnetism and thus provides an excellent playground for future investigation of low-dimensional magnetism with magnetic van der Waals materials.

**Acknowledgments**

Work at CQM and SNU was supported by the Leading Researcher Program of the National Research Foundation of Korea (Grant no. 2020R1A3B2079375). The work at the IBS CCES was supported by the Institute of Basic Science (IBS) in Korea (Grant No. IBS-R009-G1). The INS experiment was performed at the MLF of J-PARC under a user program (Proposal Nos.2019S0102).

[1] V. L. Berezinskii, Sov. Phys. JETP **59**, 907 (1971).

[2] J. M. kosterlitz and D. J. Thouless, J. Phys. C Solid State Phys. **6**, 1181 (1973).

[3] N. D. Mermin and H. Wagner, Phys. Rev. Lett. **17**, 1133 (1966).

[4] J. G. Park, J. Phys. Condens. Matter **28**, 301001 (2016).

[5] R. Brec, Solid State Ionics **22**, 3 (1986).

[6] C. Starr, F. Bitter, and A. R. Kaufmann, Phys. Rev. **58**, 977 (1940).

[7] I. Tsubokawa, J. Phys. Soc. Japan **15**, 1664 (1960).

[8] B. Huang, G. Clark, E. Navarro-Moratalla, D. R. Klein, R. Cheng, K. L. Seyler, Di. Zhong, E. Schmidgall, M. A. McGuire, D. H. Cobden, W. Yao, D. Xiao, P. Jarillo-Herrero, and X. Xu, Nature **546**, 270 (2017).

[9] C. Gong, L. Li, Z. Li, H. Ji, A. Stern, Y. Xia, T. Cao, W. Bao, C. Wang, Y. Wang, Z. Q. Qiu, R. J. Cava, S. G. Louie, J. Xia, and X. Zhang, Nature **546**, 265 (2017).

[10] S. Son, M. J. Coak, N. Lee, J. Kim, T. Y. Kim, H. Hamidov, H. Cho, C. Liu, D. M. Jarvis, P. A. C. Brown, J. H. Kim, C. H. Park, D. I. Khomskii, S. S. Saxena, and J. G. Park, Phys. Rev. B Rapid Comm. **99**, 041402(R) (2019).

[11] M. A. McGuire, Crystals **7**, 121 (2017).

[12] K. S. Burch, D. Mandrus, and J. G. Park, Nature **563**, 47 (2018).

[13] G. Ouvrard, R. Brec, and J. Rouxel, Mater. Res. Bull. **20**, 1181 (1985).

[14] C. T. Kuo, M. Neumann, K. Balamurugan, H. J. Park, S. Kang, H. W. Shiu, J. H. Kang, B. H. Hong, M. Han, T. W. Noh, and J. G. Park, Sci. Rep. **6**, 20904 (2016).

[15] S. Lee, K. Y. Choi, S. Lee, B. H. Park, and J. G. Park, APL Mater. **4**, 086108 (2016).

[16] P. A. Joy and S. Vasudevan, Phys. Rev. B **46**, 5425 (1992).

[17] D. Lançon, H. C. Walker, E. Ressouche, B. Ouladdiaf, K. C. Rule, G. J. McIntyre, T. J. Hicks, H. M. Rønnow, and A. R. Wildes, Phys. Rev. B **94**, 214407 (2016).

[18] A. R. Wildes, K. C. Rule, R. I. Bewley, M. Enderle, and T. J. Hicks, J. Phys. Condens. Matter **24**, 416004 (2012).

[19] A. R. Wildes, B. Roessli, B. Lebech, and K. W. Godfrey, J. Phys. Condens. Matter **10**, 6417 (1998).

[20] A. R. Wildes, H. M. Rønnow, B. Roessli, M. J. Harris, and K. W. Godfrey, Phys. Rev. B - Condens. Matter Mater. Phys. **74**, 094422 (2006).

[21] D. Lançon, R. A. Ewings, T. Guidi, F. Formisano, and A. R. Wildes, Phys. Rev. B **98**, 134414 (2018).

[22] N. Chandrasekharan and S. Vasudevan, J. Phys. Condens. Matter **6**, 4569 (1994).

[23] K. Kim, S. Y. Lim, J. U. Lee, S. Lee, T. Y. Kim, K. Park, G. S. Jeon, C. H. Park, J. G. Park, and H. Cheong, Nat. Commun. **10**, 345 (2019).

[24] A. R. Wildes, V. Simonet, E. Ressouche, R. Ballou, and G. J. McIntyre, J. Phys. Condens. Matter **29**, 455801 (2017).

[25] J. U. Lee, S. Lee, J. H. Ryoo, S. Kang, T. Y. Kim, P. Kim, C. H. Park, J. G. Park, and H. Cheong, Nano Lett. **16**, 7433 (2016).

[26] A. R. Wildes, V. Simonet, E. Ressouche, G. J. McIntyre, M. Avdeev, E. Suard, S. A. J. Kimber, D. Lançon, G. Pepe, B. Moubaraki, and T. J. Hicks, Phys. Rev. B - Condens. Matter Mater. Phys. **92**, 224408 (2015).

[27] K. Kim, S. Y. Lim, J. Kim, J. U. Lee, S. Lee, P. Kim, K. Park, S. Son, C. H. Park, J. G. Park, and H. Cheong, 2D Mater. **6**, 041001 (2019).

[28] G. Jackeli and G. Khaliullin, Phys. Rev. Lett. **102**, 2 (2009).

[29] R. Sano, Y. Kato, and Y. Motome, Phys. Rev. B **97**, 1 (2018).

[30] K. Tomiyasu, M. K. Crawford, D. T. Adroja, P. Manuel, A. Tominaga, S. Hara, H. Sato, T. Watanabe, S. I. Ikeda, J. W. Lynn, K. Iwasa, and K. Yamada, Phys. Rev. B **84**, 054405 (2011).

[31] F. Wallington, A. M. Arevalo-Lopez, J. W. Taylor, J. R. Stewart, V. Garcia-Sakai, J. P. Attfield, and C. Stock, Phys. Rev. B - Condens. Matter Mater. Phys. **92**, 125116 (2015).

[32] K. A. Ross, J. M. Brown, R. J. Cava, J. W. Krizan, S. E. Nagler, J. A. Rodriguez-Rivera, and M. B. Stone, Phys. Rev. B **95**, 144414 (2017).

[33] S. Itoh, T. Yokoo, D. Kawana, H. Yoshizawa, T. Masuda, M. Soda, T. J. Sato, S. Satoh, M. Sakaguchi, and S. Muto, J. Phys. Soc. Japan **82**, SA033 (2013).

[34] R. T. Azuah, L. R. Kneller, Y. Qiu, P. L. W. Tregenna-Piggott, C. M. Brown, J. R. D. Copley, and R. M. Dimeo, J. Res. Natl. Inst. Stand. Technol. **114**, 341 (2009).

[35] S. Toth and B. Lake, J. Phys. Condens. Matter **27**, 166002 (2015).

[36] J. Kennedy and R. Eberhart, in *Proc. ICNN'95 - Int. Conf. Neural Networks, Vol 4* (1995), pp. 1942–1948.

[37] M. E. Lines, Phys. Rev. **131**, 546 (1963).

[38] L. P. Regnault, C. Boullier, and J. E. Lorenzo, Heliyon **4**, e00507 (2018).

[39] H. S. Nair, J. M. Brown, E. Coldren, G. Hester, M. P. Gelfand, A. Podlesnyak, Q. Huang, and K. A. Ross, Phys. Rev. B **97**, 134409 (2018).

[40] B. L. Chittari, Y. Park, D. Lee, M. Han, A. H. Macdonald, E. Hwang, and J. Jung, Phys. Rev. B **94**, 184428 (2016).

[41] J. B. Fouet, P. Sindzingre, and C. Lhuillier, Eur. Phys. J. B **254**, 241 (2001).

[42] S. Y. Kim, T. Y. Kim, L. J. Sandilands, S. Sinn, M. C. Lee, J. Son, S. Lee, K. Y. Choi, W. Kim, B. G. Park, C. Jeon, H. Do Kim, C. H. Park, J. G. Park, S. J. Moon, and T. W. Noh, Phys. Rev. Lett. **120**, 136402 (2018).

[43] K. Takubo, T. Mizokawa, J. Y. Son, Y. Nambu, S. Nakatsuji, and Y. Maeno, Phys. Rev. Lett. **99**, 037203 (2007).

[44] K. Momma and F. Izumi, J. Appl. Crystallogr. **44**, 1272 (2011).

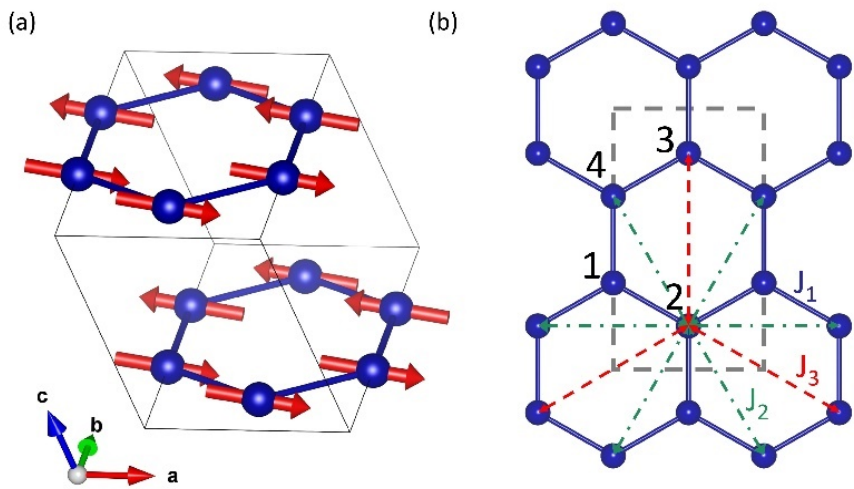

**Fig.1** (a) The magnetic structure of CoPS$_3$ with the crystallographic unit cell drawn using VESTA [44]. (b) The magnetic exchange paths are shown in the *ab*-plane for the first, second, and third nearest neighbors. The numbers denote the position of spins used for the spin-wave calculation.

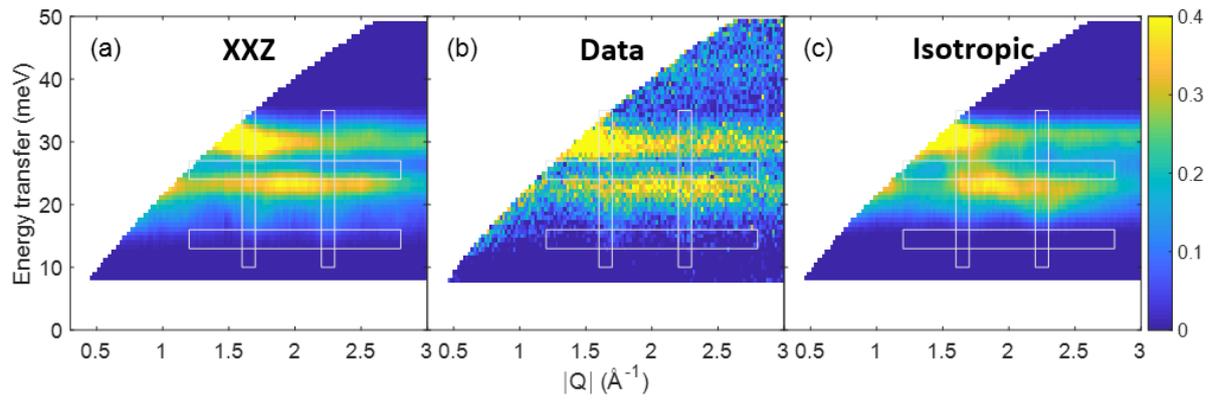

**Fig. 2.** (a) The best-fit magnon spectra with the *XXZ* model. (b) The experimental INS data of CoPS$_3$ measured at *T*=8 K with $E_i$=71.3 meV. (c) The best fit magnon spectra with the isotropic Heisenberg model. Instrumental energy resolution of 3 meV was used to convolute the theoretical results shown in (a) and (c). Horizontal and vertical white boxes denote the integration range for the constant-*E* and constant-*Q* cuts in Fig. 3, respectively.

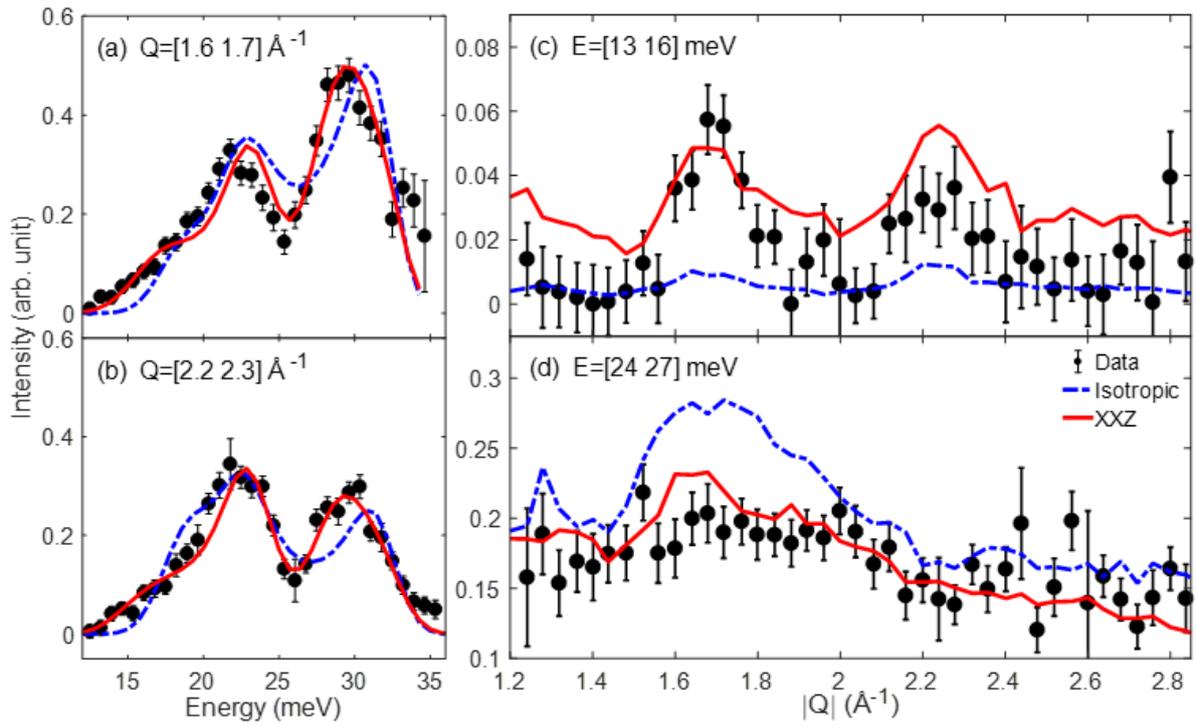

**Fig.3** (a,b) Constant-$Q$ cut at the momentum range of $Q$=[1.6 1.7] and $Q$=[2.2 2.3] Å$^{-1}$ for the measured data with the best fit simulations. (c,d) Constant-$E$ cut with the energy range of $E$=[13 16] and $E$=[24 27] meV. The *XXZ* model (solid red line) agrees better with the data.

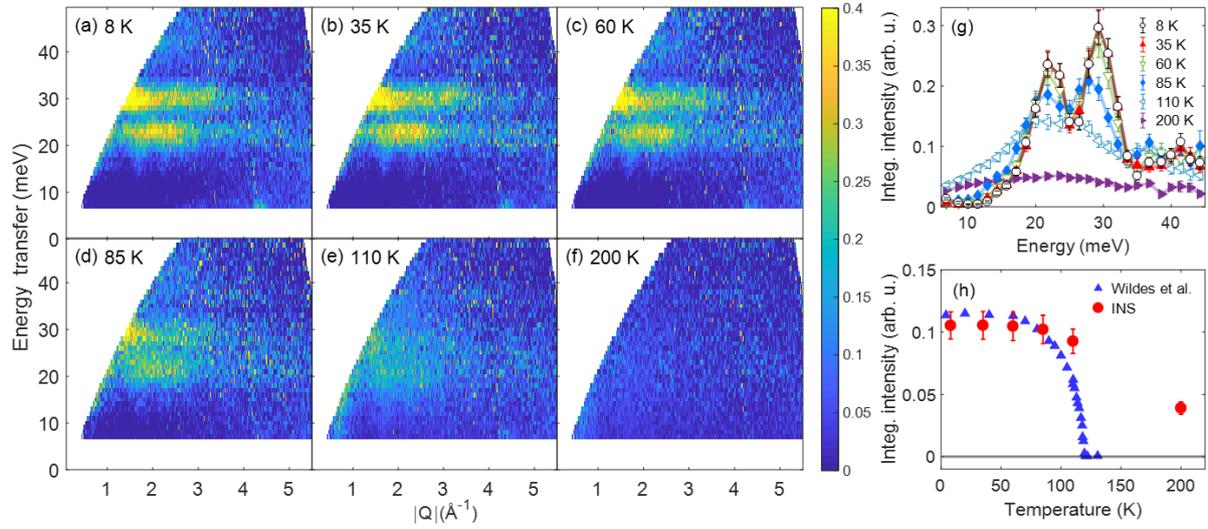

**Fig.4** (a-f) Magnon spectra at different temperatures. (g) Temperature dependence of magnon intensity integrated over the momentum range of $Q$=[0.3, 4] Å$^{-1}$ (h) Temperature dependence of the overall integrated magnon spectra. The blue triangle shows the integrated intensity of (0,1,0) magnetic peak in Ref. [24], and the red circle shows our data. The reference data were scaled to compare with our data directly.

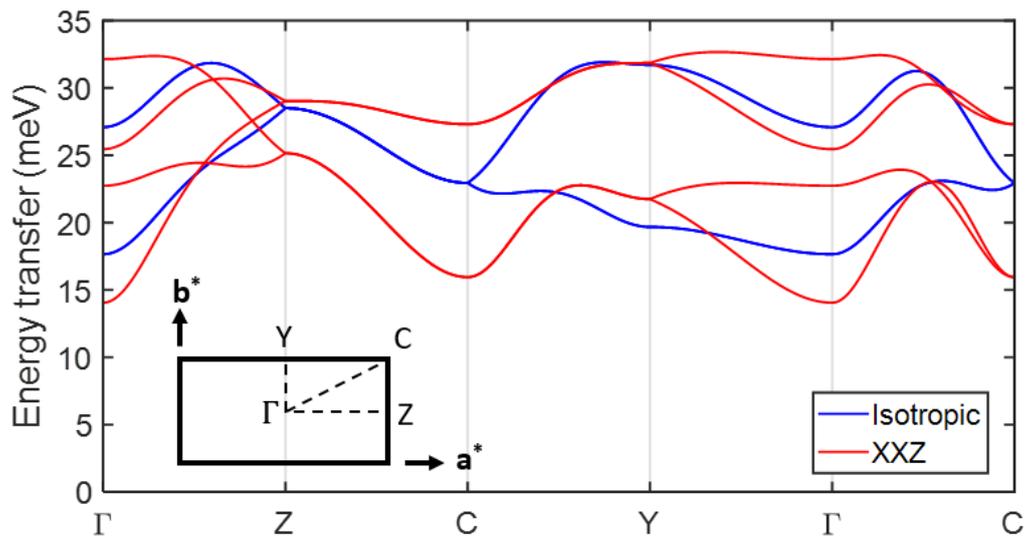

Fig 5. The spin-wave dispersion along with the high-symmetric points in the Brillouin zone for the isotropic Heisenberg model and the *XXZ* model. All the trajectories are given for the crystallographic unit cell. The Brillouin zone and the relevant positions are shown in the insert.

**Table 1.** Magnetic exchange and anisotropy parameters for the magnetic vdW *TM*PS$_3$ (*TM*=Mn, Fe, Co, and Ni) with the parameters defined in a self-consistent way with Eq. (1).

|  | MnPS$_3$ [19] | FePS$_3$ [17] | CoPS$_3$ (this work) | NiPS$_3$ [21] |
|---|---|---|---|---|
| $S$ | 5/2 | 2 | 3/2 | 1 |
| $T_N$ (K) | 78 | 120 | 120 | 155 |
| $J_1$ (meV) | 1.54 | -2.96 | -2.04 | -3.8 |
| $J_2$ (meV) | 0.14 | 0.08 | -0.26 | 0.1 |
| $J_3$ (meV) | 0.36 | 1.92 | 4.21 | 13.8 |
| α | 1 | ∞ | 0.6 | 1 |
| $K$ (meV) | -0.0086 | -2.66 | -2.06 | -0.3 |